\documentclass[prd,letterpaper,onecolumn,amsmath,amsfonts,amssymb,nofootinbib]{revtex4-2}
\pdfoutput=1
\usepackage {amsmath}  
\usepackage[dvips]{graphicx}                      
\usepackage{float}           
\usepackage{feynmp}         
\usepackage{amssymb}                                
\usepackage[colorlinks=true                                                                             
,urlcolor=blue                
,anchorcolor=blue                      
,citecolor=blue         
,filecolor=blue            
,linkcolor=blue      
,menucolor=blue       
,pagecolor=blue    
,linktocpage=true. 
,pdfproducer=medialab                              
,pdfa=true            
]{hyperref}  
 \usepackage{csquotes}      
\newcommand{\be}{\begin{equation}}
\newcommand{\ee}{\end{equation}}
\newcommand{\bear}{\begin{eqnarray}} 
\newcommand{\eear}{\end{eqnarray}}

\newcommand{\vev}[1]{\left\langle #1\right\rangle}

\newcommand{\lapproxeq}{\lower .7ex\hbox{$\;\stackrel{\textstyle
<}{\sim}\;$}}
\newcommand{\gapproxeq}{\lower .7ex\hbox{$\;\stackrel{\textstyle
>}{\sim}\;$}}
\newcommand{\stackdown}[2]{\lower 1.4ex\hbox{$\;\stackrel{\textstyle{#1}}
{\scriptstyle{#2}}\;$}}
\newcommand{\beq}{\begin{equation}}
\newcommand{\eeq}{\end{equation}}

\newcommand{\ba}{\begin{eqnarray}}
\newcommand{\ea}{\end{eqnarray}}

\newcommand{\bea}{\begin{eqnarray}}
\newcommand{\eea}{\end{eqnarray}}

\makeatletter
\def\slash{\@ifnextchar[{\fmsl@sh}{\fmsl@sh[0mu]}}
\def\fmsl@sh[#1]#2{%
  \mathchoice
    {\@fmsl@sh\displaystyle{#1}{#2}}%
    {\@fmsl@sh\textstyle{#1}{#2}}%
    {\@fmsl@sh\scriptstyle{#1}{#2}}%
    {\@fmsl@sh\scriptscriptstyle{#1}{#2}}}
\def\@fmsl@sh#1#2#3{\m@th\ooalign{$\hfil#1\mkern#2/\hfil$\crcr$#1#3$}}
\makeatother
\usepackage{color}

\definecolor{orange}{rgb}{0.9,0.2,0}
\definecolor{brown}{rgb}{0.7,0.3,0.2}
\definecolor{fuxia}{rgb}{1,0,1}
\definecolor{skyblue}{rgb}{0,0.1,0.9}
\definecolor{violetred}{rgb}{0.8,0.13,0.56}
\definecolor{deeppink}{rgb}{1.00,0.08,0.5}
\definecolor{pink}{rgb}{1.00,0.75,0.80}
\definecolor{orchid}{rgb}{0.85,0.44,0.84}
\definecolor{lightpink}{rgb}{1.00,0.71,0.76}
\definecolor{bluish}{rgb}{0,0.6,0.8}  
\begin{document}
\title{Aspects of supersymmetry breaking driven inflation in orbifold models. }
\author{
\vspace*{8mm}
{\bf G. A.~\ Diamandis} {\footnote{email: gdiam@phys.uoa.gr}}, \,  {\bf K.~\ Kaskavelis}{\footnote{email: kkaskavelis@phys.uoa.gr}}, , \,  {\bf A. B.~ \ Lahanas}{\footnote{email: alahanas@phys.uoa.gr}}, \,{\bf  G.~\ Pavlopoulos}{\footnote{email: gepavlo@phys.uoa.gr}}}
\affiliation{National and Kapodistrian University of Athens, Department of Physics,\\
Nuclear and Particle Physics Section, GR--157 71  Athens, Greece}

\vspace*{1cm}    
  
\begin{abstract}
We consider gravitationally induced corrections to inflaton potentials driven by supersymmetry breaking in a five-dimensional supergravity, compactified on a $ S_1/Z_2 $ orbifold.  The supersymmetry breaking takes place on the hidden brane and is transmitted to the visible brane through finite one loop graphs giving rise to an inflaton potential which includes  
 gravitationally induced terms.  These corrections are significant for inflationary cosmology and have the potential to modify the predictions of widely studied supergravity models if the latter are embedded in this framework.  To explore these effects we examine two classes of models those inspired by  no-scale supergravity models and  $\alpha$-attractors. Both models are compatible  with current cosmological observations but face chalenges in reconciling enhanced values for the scalar power spectrum $ P_\zeta$ with cosmological data, particularly regarding the tensor to scalar ratio $r$.  In fact  $ P_\zeta \gtrsim 10^{-2}$ results to 
 $ r > \mathcal{O} (0.1) $, outside the limits put by current data. 
\end{abstract} 
\maketitle
 

\section{Introduction} 

It is known that $N=2$, $d=5$, supergravity \cite{deWit:1982bul, 
Pernici:1985ju, Gunaydin:1983bi, Gunaydin:1983rk, 
 Gunaydin:1984pf, deWit:1991nm, Gunaydin:1999zx, Gunaydin:2000xk, Ellis:1999kd, Falkowski:2000yq, Bergshoeff:2000zn}, compactified on $S_1/Z_2$ orbifolds leads to $N=1$ local supersymmetry  localized on two branes one of which may be considered to be the visible brane the other being the hidden one. The $N=1$ local supersymmetry is implemented by the fields even under $Z_2$ that is the graviton multiplet and the radion multiplet. Note that the real part of the scalar field of the radion multiplet is $e_{5}^{\,\dot{5}}$, the $(5, \, \dot{5})$ component of the f\"{u}nfbein.  

The couplings of chiral multiplets,  localized on  the visible brane,  to the radion multiplet is achieved by extending   the K\"{a}hler  function $\, K( \phi_i, \phi^*_i) $ to \cite{Diamandis:2004uk}
\be
{\cal{F}} \, = \, -3 \, ln \, \frac{T+T^*}{\sqrt{2}} \, + \, \Delta_{(5)} \,  K( \phi_i, \phi^*_i )\,,
\label{branepast}
\ee
where $ \Delta_{(5)} \,=\, \frac{\sqrt{2}}{T+T^*} \delta (x_5) $ and 
\be
\frac{T + T^*}{\sqrt{2}}\, =\,  e_5^{\,\dot{5}}, \quad \frac{\sqrt{2}}{T + T^*}\, =\,  e_{\dot{5}}^{\,5}\,.
\label{radion}
\ee

Using ${\cal{F}}$ the part of the Lagrangian describing the kinetic terms are obtained as
\be
e^{(5)} {\cal{L}}_{Kin} \,=\, -e^{(5)}\,{\cal{F}}_{ab^*} \partial_{\mu} \phi^a \partial^{\mu} \phi^{b \, *}  \, , 
\label{kinorig}
\ee
where $a = T, i$.
With the expression above we derive  the kinetic terms of the fields on the brane including the restriction of the radion field on it and the relevant derivative terms describing the coupling of the brane fields to the radion at the kinetic part.
Moreover given a superpotential $ W(\phi_i) $ on the visible brane the arising potential has been shown to be \cite{Diamandis:2004uk},
\be
e^{(5)} {\cal{L}}_P \,=\, -e^{(5)}\,\left( \Delta_{(5)} \right)^2 \,e^{\cal{F}} \left( {\cal{F}}^{a b^*} D_a W D_{b^*} W^* - 3 \Delta_{(5)} W W^*   \right)\,. 
\label{potorig}
\ee
 Chiral multiplets on the hidden brane are treated analogously. 

It is obvious that the whole  Lagrangian is derived as an expansion in powers of $\Delta_{(5)}$. The terms linear in $\Delta_{(5)}$ yield the first approximation of the brane action while the divergent higher powers are significant dealing with quantum corrections.

In this work we are interested in exploring cosmological features of models adopting this point of view. For this purpose, we consider only the terms linear in $\Delta_{(5)}$ and fix the radion field at a constant value. In doing so we adopt the first approximation in the brane-bulk coupling and we do not address the issue of radion stabilization. 
In this framework the  kinetic term on the brane is
 \be
- K_{ij^*} \partial^\mu \phi^i  \partial_\mu \phi^{j \,*}\,,
 \ee
 and the potential of Eq.(\ref{potorig}) becomes
\be
 \left(  \frac{ \sqrt{2}} { T+T^*}  \right)^{ \frac{3}{2}} \,      K^{ij^*} 
 \frac{\partial W}{\partial \phi_i} \frac{\partial W^*}{\partial \phi_j^*}\,.
 \ee
So in this approximation the effect of the radion field is to give back the flat case potential, eliminating the negative contribution in Eq.(\ref{potorig})  and rendering the covariant derivatives to simple ones.  Note the appearance of the prefactor   that depends on the radion field  $T$.  Assuming that the value of $T$ is fixed to be a  constant these terms can be absorbed within the couplings of the superpotential.  This will be explored further in the context of specific  models discussed in the sequel.

One of the most interesting characteristics of these constructions is that the supersymmetry breaking may occur just by adding a constant superpotential on the hidden brane \cite{Mirabelli:1997aj, Bagger:2001qi, Bagger:2001ep, Falkowski:2000er, Gherghetta:2001sa, Gherghetta:2001sa}. The supersymmetry breaking is transmitted to the visible brane by finite one loop diagrams with vertices on both branes leading to the "soft" supersymmetry breaking terms \cite{Rattazzi:2003rj, Scrucca:2004cw, Diamandis:2007zq}
\be
m_{\phi}^2 \,K + 3 \frac{m_{\phi}^2}{m_{3/2}} \left( W + c.c. \right)\,,
\label{softterms}
\ee 
where
\be
m_{\phi}^2 \,=\, \frac{k_5^2}{16} \frac{\zeta(3)}{\pi^5 R^3} \,  m_{3/2}^2     \,=\,\frac{\zeta(3)}{8 \pi^4 R^2} \frac{m_{3/2}^2}{m_{Planck}^2}, \quad m_{3/2} \,=\, \frac{k_5^2 |w_0|}{\pi R}\,,
\label{branepar}
\ee
with $k_5$ being  the five dimensional gravitational coupling constant and   $m_{Planck} $ the reduced Planck mass. 
$R$ is the distance between the two branes and $w_0$  the  superpotential on the hidden brane, which for simplicity is assumed constant. Note that in the case of canonical kinetic terms, i.e.  
$\, K( \phi, \phi^*) \,=\, \phi^{*}\phi $,  the first term in Eq.(\ref{softterms}) is just the shift in the mass of the scalar fields with respect to their spinor counterparts of the chiral multiplets.

Before proceeding further,  to  discuss  in detail  specific models,  we should comment on a basic characteristic of  this framework. Considering the simplest possible supersymmetric case described by the K\"{a}hler  function $\, K( \phi, \phi^*) \,=\, \phi^{*}\phi $ leading to canonical kinetic terms and a cubic superpotential $ W(\phi) \propto \phi^3$ we see that the resulting supersymmetric potential is $V_0 \,\propto \left( \phi^{*} \phi  \right)^2$. The  inflationary behavior from such potential is a genuine two-field case. Two-field inflation with canonical kinetic terms is not attractive,  if not excluded \cite{Peterson:2011yt}, although models with noncanonical kinetic terms are very interesting \cite{Peterson:2011yt, Greenwood:2012aj, Kaiser:2013sna, Christodoulidis:2019jsx, Boutivas:2022qtl}. The inclusion of the terms stemming  from the supersymmetry breaking in Eq.({\ref{softterms}}) makes the cosmological evolution to be effectively that of one-field, the real part of the field $\phi$. In this case we have inflationary behavior which may exhibit, under certain conditions,  
inflection points necessary for  an Ultra Slow-Roll  (USR) phase during which the 
first Hubble-flow function $\epsilon_1 = - \frac{\dot{H}}{  H^2} $ is of order unity. This may  trigger an augmented enough power spectrum for 
 wavelenghts smaller than CMB length scales.  Whether this is adequate  to implement a strong enough USR phase, pertinent to production of primordial black holes \cite{Carr:1974nx, Meszaros:1974tb, Carr:1975qj, 1979A&A....80..104N, Kinney:2005vj, Ballesteros:2017fsr, Sasaki:2018dmp, Dalianis:2021iig, Carr2024historyprimordialblackholes, choudhury2024largefluctuationsprimordialblack},  
 and be compatible with cosmological data,  will be discussed when considering particular models. 


\section{No-scale Model}

In this section we shall consider a model whose   K\"{a}hler function is reminiscent of the   no-scale supergravity \cite{Cremmer:1983bf, Ellis:1983sf}  which arises  as an effective description of a more fundamental theory. In fact we shall adopt 
\[
K \,=\, -3 M^2 ln\left( 1 - \frac{\phi \phi^*}{3 M^2}   \right) \,, 
\]
assuming a   superpotential  of the form 
\[
W \,=\, \frac{\lambda^\prime}{6} \phi^3 \,+\, \frac{m^\prime}{2} \phi^2, \quad \text{with}\,\,\lambda^\prime, m^\prime \,\,\text{real} \,.
\]
We are working in Planck unites, $m_{Planck} = 1$,  and all quantities appearing above are dimensionless. 
The case  $ M = 1 $ corresponds to the well-known no-scale  K\"{a}hler potential but we have also allowed for  other values, $M \neq 1$.  

In this case the scalar field kinetic term are given by, 
\[
-\frac{1}{\left(1 - \frac{\phi \phi^*}{3 M^2}\right)^2} \partial_\mu \phi \partial^\mu \phi^* \, 
\]
while the   potential on the brane stemming  from the superpotential $W$ is
\be
V_{brane} \,=\, \left( 1 - \frac{\phi \phi^*}{3 M^2} \right)^2 \left[ \frac{\lambda^2}{4}(\phi \phi^*)^2 + \frac{\lambda m}{2}\phi \phi^*(\phi + \phi^*) + m^2 \phi \phi^* \right] \,.
\label{noscpot}
\ee
Note that we have redefined the original superpotential couplings, to absorb the effect of the radion field,  
\bea
\lambda^\prime = c \, \lambda  \quad , \quad  m^\prime = c \, m \quad \text{with}  \quad c = {\vev{  \frac{ T+T^*}{\sqrt{2} } } }^{3/2}\, .
\eea
The contribution from the transmission of the supersymmetry breaking reads
\begin{eqnarray}
&& m_{\phi}^2\,K \,+\,\frac{3 m_{\phi}^2}{m_{3/2}} \,\left( W \,+\, W^*  \right)    \nonumber \\
&&= \,  -3 M^2 m_{\phi}^2 ln \left(1 - \frac{\phi \phi^*}{3 M^2}\right) \,+\, 
\frac{3 m_{\phi}^2 c}{m_{3/2}}   \left[\dfrac{\lambda}{6} (\phi^3 + \phi^{*3}) + \dfrac{m}{2} (\phi^2 + \phi^{*2})  \right]\,.
\end{eqnarray}
Now if we consider the direction along the real part of the scalar field  $Re[\phi] \,=\, x/\sqrt{2}$ we get for the full potential
\[
V(x)  \,=\,  \left( 1 - \frac{x^2}{6M^2}  \right)^2 \left[ \frac{\lambda^2}{16} x^4 + \frac{\lambda m}{2 \sqrt{2}}x^3 +\frac{ m^2 }{2} x^2  \right] \,-\, 3 M^2 m_{\phi}^2 ln \left(1 - \frac{x^2}{6 M^2}\right)\,+\, \frac{3 m_{\phi}^2}{ m_{3/2}} \left[ \frac{\lambda}{6 \sqrt{2}} x^3 + \frac{m}{2}x^2 \right]\,.
\]
Besides the trasformation
\[
x \,=\, \sqrt{6}M \, tanh\frac{\varphi}{\sqrt{6}M}\,, 
\]
gives canonical kinetic terms for the field $\varphi$.
Summarizing we see that in terms of the canonically normalized field the overall potential takes the form
\be
V(\varphi) \,=\, V_0 \left[ \frac{tanh^2 f}{cosh^4 f} \left(2 l_2 + tanh f  \right)^2 \,+\, m_1 \,ln \left( cosh^2 f \right) \,+\, m_3\, tanh ^3 f \,+\, 3 m_3\, l_2\, tanh^2 f
\right]\,,
\label{fullpot}
\ee
where $f \,=\, \varphi /(\sqrt{6}M)$ and 
\be
V_0 \,=\, \frac{9 \lambda^2 M^4}{4}, \,\,\, l_2 \,=\, \frac{m}{\sqrt{3} \lambda M},\,\,\,m_1\,=\,\frac{4}{3} \frac{m_{\phi}^2}{\lambda^2 M^2}, \,\,\, m_3 \,=\, 
\frac{4}{\sqrt{3}}\frac{ m_{\phi}^2}{ m_{3/2}} \frac{c}{\lambda M}\,.
\label{parameters}
\ee
It becomes evident that, in the absence of  supersymmetry breaking  in Eq.(\ref{fullpot}) we do not have any interesting cosmological behavior since the potential is positive definite tending to zero as $f \rightarrow \pm \infty $ unlike the simple four dimensional case \cite{Ellis:2015xna, Ellis_2020}.  If we consider the supersymmetry breaking the relative basic parameters  in this setting are given by
\be
m_{\phi}^2 \,=\, \frac{m_1}{3} \frac{V_0}{M^2}, \quad m_{3/2} \,=\,\frac{2 m_1}{\sqrt{3} |m_3|} \frac{\sqrt{V_0}}{M}  c\,.
\label{susypar}
\ee 
Taking the inflation scale to be   $V_0 \approx 10^{-10} $, which is close to that dictated by COBE normalization,  we derive  
\[
m_{\phi} \, , \, m_{3/2}   \approx \dfrac{10^{-5}}{M \,}  
\]
in Planck units.  Note that we have assumed $c \approx 1 $ in deriving the value of the  gravitino mass. The above relations state that 
for $M=1$ we have masses of the order of  $10^{-5} $, in Planck units,   equivalent to $10^{13} GeV$. It is interesting to see that in order to have lower  masses   $M > 1 $ is needed,  equivalent to  $M > m_{Planck} $ in GeV units, i.e. we need  transplanckian masses. 

The cosmological predictions of these  models have to be compared  with the latest cosmological data by various sources,  The spectral scalar index $\eta_s$ 
should be within  $\eta_s = 0.9649 \pm 0.0044 $   ( $Planck \, TT, TE, EE + lowE + lensing + BICEP2/Keck Array $)  or 
$\eta_s = 0.9668 \pm 0.0036 $  if the BAO data are also combimed  \cite{Planck:2018jri}, 
 while the tensor-to-scalar ratio $r$ should be bounded from above $ r < 0.063 $ if Planck and BICEP/Keck Array data are used ( $Planck \, TT, TE, EE + lowE + lensing + BICEP+BAO $) \cite{Planck:2018jri}.  For the bound of $r$ the consistency relation $\eta_T = -r/8 $ has been used, for the tensor tilt, which is the case for a single field  slow-roll inflation. Omitting the data set for lensing and Baryon Accoustic Oscillations (BAO) the value of $r$ may be larger 
$r < 0.11$. More recent studies however tend to lower considerably the upper bound on $r$ yielding $ r < 0.032$, 
\cite{Tristram:2021tvh,BICEP:2021xfz,eBOSS:2020yzd,Planck:2018lbu}. Other studies \cite{Campeti:2022vom}, yield slightly larger upper bounds on $r < 0.036 $ or so. 

For the model at hand,  taking for instance the parameters $l_2 = 2.0, m_1 = 0.5, m_3 = 1.0$ and the scale $V_0 = 1.3 \times 10^{-10}$, the predictions are, 
\bea
\eta_s = 0.9669 \,,\quad r = 0.035 \,.
\nonumber
\eea
The choice of the scale $V_0$ is adjusted so that the the spectrum corresponding to the CMB scale $k = 0.05 \, Mpc^{-1}$ is within observational limits ( COBE normalization ). For the chosen value of $V_0$  we have
\bea
P_{0.05} =  2.13 \times 10^{-9} \,.
\eea

For  the previous choice of the parameters the scalar power spectrum exhibits a smooth behaviour for wave numbers ranging from low to large values, which is smaller for larger 
$k$. It would be interesting to examine if this type of models can sustain cases where an enhanced power spectrum can be  predicted relevant for production of primordial black holes \cite{Nanopoulos:2020nnh}. Towards this goal an inflection point should be developed giving rise to an USR ( Ultra Slow Roll ) phase during the inflationary epoch. 
In studying the cosmological evolution for the inflaton field $\varphi$ we see that we can find an inflection point satisfying 
\[
V^{''}(\varphi_{in}) \,=\, 0, \quad V^{'}(\varphi_{in}) \,\approx \,0, \quad \text{with} \quad   \delta \equiv \frac{V^{'}(\varphi_{in})}{V(\varphi_{in})}   \approx 0
\]
and consequently a USR phase in the vicinity of this inflection point may emerge.

For values of the  parameters giving rise to a USR phase the power spectrum can be indeed augmented, in the range  $P_\zeta \sim 10^{-4, -5}$   pertinent to the creation of primordial black holes,  but the tensor to scalar ratio and the spectral index turn out to be outside the bounds put by cosmological data, in fact we find that $r \sim 0.12  $  and $\eta_s \sim 0.960 $,  the latter being slightly lower than its observational limit.  One can obtain larger values for the power spectrum,  $P_\zeta \sim 10^{-2}$, for a proper fine tuned values of the parameters involved,  at the cost of having even large values of $r \gtrsim 0.15$  and lower  
$\eta_s \lesssim 0.943$ well outside the current limits  put by data, so we totally  disregard these cases.  As a first example, with the following choice 
 of the parameters determining the potential in Eq.(\ref{fullpot}) 
\be 
M\,=\,1,\,\,l_2 \,=\, 1.05, \,\, m_1 \,=\, 7.50638,  \,\, m_3 \,=\, -2.97221  \,,
\label{first}
\ee 
we find an inflection point at $\varphi_{in} \,=\, 1.697$ with $\delta \,=\, -0.014$. 
 For the specific choice the model predicts
\bea
r = 0.117 \quad , \quad  \eta_s = 0.956     \quad , 
\eea
while the spectrum   gets it maximum value $P_{max} \simeq 9.0 \times 10^{-5}$ at   $k_{max} \simeq 2.3 \times 10^{14} \,  Mpc^{-1}$.  
 The scale of the potential is taken $V_0 = 7.3 \times 10^{-11}$  to be compatible with    $P_{0.05} =  2.12 \times 10^{-9} \,  $,  consistent with data.
Observe  that from Eq.(\ref{parameters}) negative value of $m_3$ and positive value of $l_2$ can be obtained if both $\lambda$ and $m$ are taken  negative. This is perfectly legitimate  since only the parameter $m_1$ ought to be positive.  

Another set of values is 
\be 
M\,=\,1,\,\,l_2 \,=\, 1.190, \,\, m_1 \,=\, 8.81571,  \,\, m_3 \,=\, - 3.14683  \,,
\label{second}
\ee
with  an inflection point at $\varphi_{in} \,=\, 1.7$ with $\delta \,=\, -0.0085$.  
 With these we have, 
\bea
r = 0.102 \quad , \quad  \eta_s = 0.9620     \quad .
\eea
while the scale of the potential is  $V_0 = 4.98 \times 10^{-11}$ giving   $P_{0.05} =  2.09 \times 10^{-9} \,  $.
The value of $r$ becomes smaller in this case, tending towards a reasonable  agreement level, although still above its allowed bound,  and  $\eta_s$ touches  its lowest  allowed observational limit.  Notice however that the spectrum in this case exhibits a maximum which is  smaller than the previously considered case, actually 
$P_{max} \simeq 0.4 \times 10^{-6}$ at   $k_{max} \simeq 3.2  \times 10^{16} \,  Mpc^{-1}$.   

\begin{figure}[t]
\centering
  \centering     
 \includegraphics[width=0.49\linewidth]{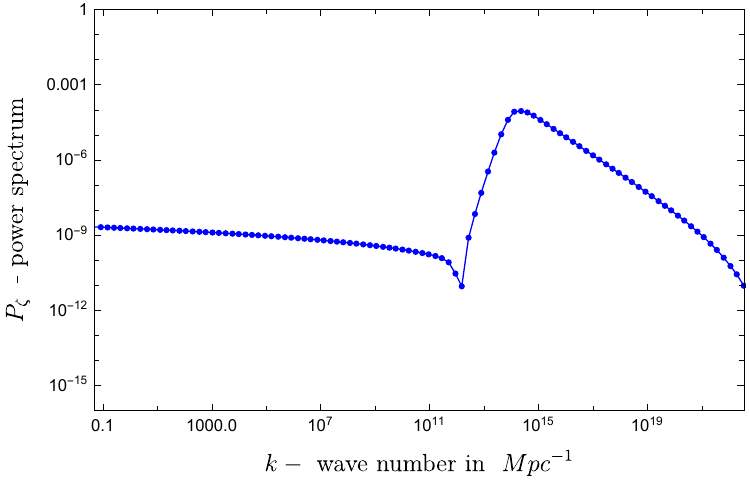}   
  \includegraphics[width=0.49\linewidth]{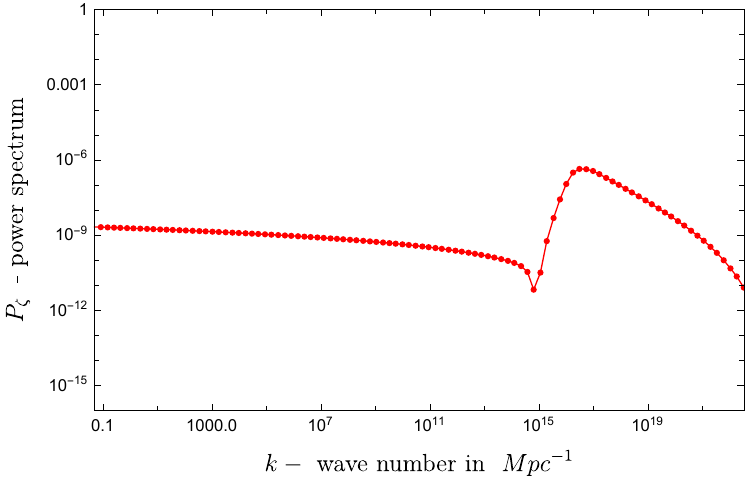}   
\caption{
The power spectrum for the no-scale  model for values of the parameters given in Eq.(\ref{first}) (left), and for the set given by Eq.(\ref{second}) (right).  For the first case (left) the spectrum exhibits a  maximum at  $P_{max} \simeq 9.0 \times 10^{-5}$ at   $k_{max} \simeq 2.3 \times 10^{14} \,  Mpc^{-1}$.  The second case (left) is more consistent with data but the spectrum is considerably lower. Actually   $P_{max} \simeq 0.4 \times 10^{-6}$ at   $k_{max} \simeq 3.2 \times 10^{16} \,  Mpc^{-1}$.  
 }
   \label{eenosca}
\end{figure}  

The conclusion concerning the cosmological predictions of no-scale models is that, complete agreement with cosmological data for   $r  \, , \,  \eta_s $ is possible,  however the power spectrum turns out to be small $ P_\zeta \lesssim 10^{-6}$, for any wave number.


\section{Attractor models}
In this section we embed the \ensuremath{\alpha}-attractor model \cite{ Kallosh:2013yoa} in our scheme. In this model the K\"{a}hler potential is
\be
K \,=\, -3  \, \alpha  \, ln \left(1 - \frac{\phi^* \phi + s^* s}{3}    \right) 
\label{dal1} 
\ee
and the superpotential is given by
\be
W \,=\, s f\left( \frac{\phi}{\sqrt{3}}  \right) \left(3 - \phi^2  \right)^{\frac{3  \alpha \, -1}{2}}\,.
\label{dal2}
\ee
 For the relevant cosmological trajectory we get $s\,=\,Im[\phi] \,=\,0$ and in order to have canonical kinetic terms we redefine  as
\[
\phi \,=\, \sqrt{3} tanh \left( \frac{\varphi}{\sqrt{6 \alpha}}  \right)\,.
\]   
In this case the potential on the brane is obtained to be
\be
V_{brane} \,=\, c^{-1} \, \frac{3^{3 \alpha-1}}{\alpha }\,cosh^{-6a}\left( \frac{\varphi}{\sqrt{6 \alpha}} \right) \, 
\left|f\left( \frac{\varphi}{\sqrt{6 \alpha}} \right)  \right|^2  \, ,
\label{dalpot}
\ee
where  the constant $c$ is defined in the previous section. 
The supersymmetry breaking adds, in this case,  the term
\be
3 \alpha \, m_{\phi}^2 \,ln \left[cosh^{2}\left( \frac{\varphi}{\sqrt{6 \alpha}} \right)  \right]\,. 
\label{dalsusy}
\ee
Note that since we work with $s = 0$ the second term in Eq.(\ref{softterms}) is absent. So in this case there is no cosmological constraint for the gravitino mass.
If we consider a  polynomial which is at most cubic for the function $f$ we have a scalar  potential
\be
V(\varphi) \,=\, \frac{3^{3 \alpha-1}}{a}\,cosh^{-6 \alpha}\left( \frac{\varphi}{\sqrt{6 \alpha}} \right) \left[ d_0 + d_1 z + d_2 z^2 + d_3 z^3  \right]^2 + 3 \alpha m_{\phi}^2 \,ln \left[cosh^{2}\left( \frac{\varphi}{\sqrt{6 \alpha}} \right)  \right]\,,
\label{daltot}
\ee 
where we have absorbed the constant $c$ of Eq.(\ref{dalpot}) within the couplings $d_i$ and the variable $z$ is given by 
$z \;=\; tanh\left( \frac{\varphi}{\sqrt{6 a}} \right) $. 
This potential exhibits a linear behaviour for high inflaton values, $ \varphi > \sqrt{6 \alpha}  $,  due to the logarithmic term which prevails over the rest of the terms in 
Eq.(\ref{daltot}).

In the following we shall assume that 
$f$ is at least quadratic in $z$, that is  $d_2 \, \neq\, 0$,  so that the potential may be cast in the  form
\be
V(\varphi) \,=\, V_0 \left\{cosh^{-6a}\left( \frac{\varphi}{\sqrt{6 \alpha}} \right) \left[ b_0 + b_1 z +  z^2 + b_3 z^3  \right]^2 + m_1 \,ln \left[cosh^{2}\left( \frac{\varphi}{\sqrt{6 \alpha}} \right)  \right]  
\right\} \,,
\label{dalfin}
\ee
where $V_0 \,=\,\frac{3^{3 \alpha-1}}{\alpha} d_2^2, \,\, b_0 \,=\, d_0/d_2, \,\, b_1 \,=\, d_1/d_2, \,\, b_3 \,=\, d_3/d_2, \,\, m_1 \,=\, \frac{3^{2 - 3 \alpha} \alpha^2 m_{\phi}^2}{d_2^2}   $.

In this model it is hard to have  low values of $r$,  irrespectively of the value of  $\eta_s$.  To be more precise, we have found that $ r \gtrsim 0.075$  and if $\eta_s$ is forced to be  within the allowed range the value of $r$ is larger, $ r \gtrsim 0.085$. This results follows by scanning the parameter space keeping $\alpha= 1 $. Similar results hold for other $\alpha$ values, as well. 

In this model too we can have cases where augmented power spectrum is observed if we fine tune the parameters of the scalar potential.  In particular for values 
$b_0 =0, b_3 = 1.2 $ and taking $b_1 = -0.938137, m_1 = 0.104430 $ an inflection point is developed at  $\varphi_{inf} =1 $, with $\delta = - 0.06882$,  and a USR phase takes place.
For these values the maximum value for the power spectrum obtained is $P_{max} \simeq 5.3 \times 10^{-4}$  in the region of 
very  high wavenumbers $k$  of the order of $k \simeq 10^{19} \, {Mpc}^{-1}$, or so.  In this example 
we have taken  the scale of the potential to be $V_0= 3.3 \times 10^{-9}$, dictated by COBE normalization,  yielding .  $P_{k=0.05} =2.1 \times 10^{-9}$ for the 
CMB scale $k = 0.05 \, {Mpc}^{-1} $.  Assuming instantaneous reheating the predictions for the spectral index and the tensor to scalar ratio corresponding to 
$ k = 0.05 \, {Mpc}^{-1} $ are as follows
\bea
\eta_s = 0.9675\,, \quad  \quad r =0.087 \, .
\eea
The value of $r$ is  outside the limits put by Planck and BICEP2/Keck array data, $ r < 0.063$, and more recent analyses, which give an even  lower upper bound,  by a factor of two or so.
  
As a second case,  we take, as before,  $b_0 =0, b_3 = 1.2 $ but we consider different values for the remaining parameters, $b_1= - 1.120235, m_1 = 0.149537 $.  In this case the inflection point is slightly shifted to the right, i.e. $\varphi_{inf} =1.1 $, with $\delta = -0.0052$. In this case higher values are obtained   $P_{max} \simeq 2.94 \times 10^{-3}$  for  smaller 
$k$-values, actually   $k_{max}  \simeq 1.95 \times 10^{13} \, {Mpc}^{-1}$. 
Note that in this case COBE normalization demands that   $V_0$ is slightly higher,  $V_0= 4.0 \times 10^{-9}$.  
With  instantaneous reheating assumed the predictions for the spectral index and the tensor to scalar ratio, for the CMB scale 
$k = 0.05 \, {Mpc}^{-1} $,  are as follows 
\bea
\eta_s = 0.9536\,, \quad  \quad r =0.125  \, .
\eea
Therefore we can have higher values for the spectrum, which may be cosmologically interesting and  relevant for production of primordial  black holes,  but the tensor to scalar ratio has moved to a region that is disfavored by current data.  This seems to be a general characteristic of these models. High values for the power spectra
$ P_\zeta > 10^{-3}$ do not reconcile with values of $r$ below $0.08$ . This is a rather general feature in these models due to their linear behaviour for high inflaton values, in conjunction with 
the presence of the $ cosh^{-6a} $ term in the potential  (\ref{daltot}) which has a drastic effect. 
\begin{figure}[t]
\centering
  \centering     
 \includegraphics[width=0.49\linewidth]{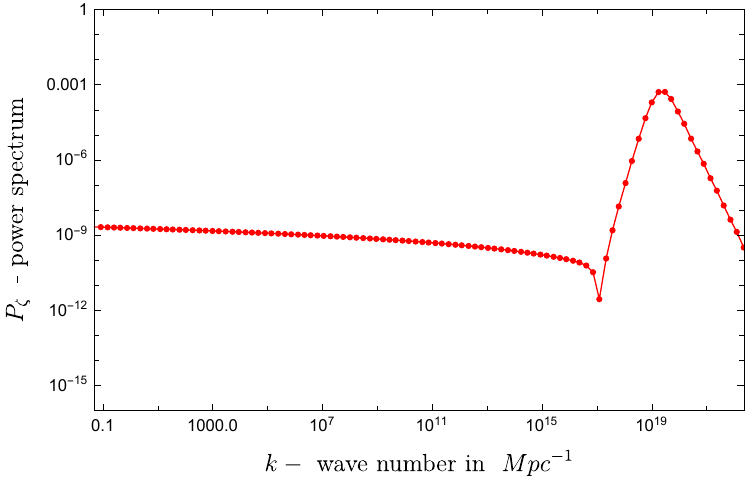}   
   \includegraphics[width=0.49\linewidth]{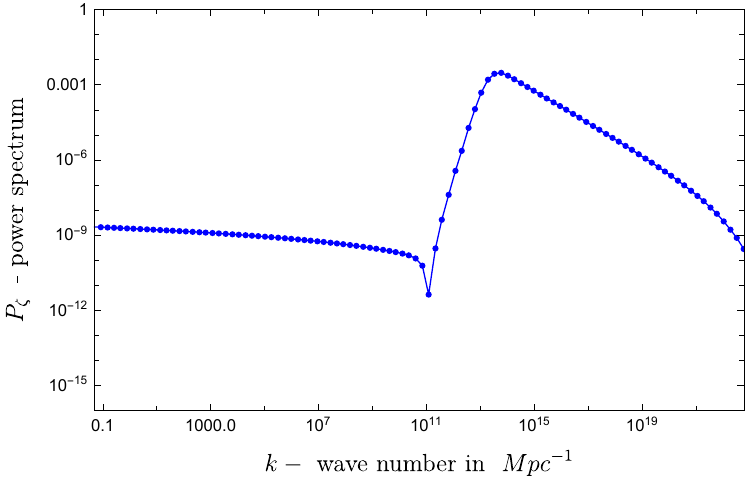}     
\caption{
The power spectrum for the $\alpha$-attractor model for values of the parameters $\alpha =1$ and $b_0=0, b_3=1.2$ and  
$ b_1= - 0.938137, m_1 = 0.104430$ ( left panel ) and  $ b_1= - 1.120235, m_1 = 0.149537$ ( rightpanel ).
For the first case the inflection point is at $\varphi_{inf} =1 $ while for the second at  $\varphi_{inf} =1.1 $ .  
 }
   \label{ee12}
\end{figure}  

To ameliorate the effect of the offending $ cosh^{-6a} $ term, discussed previously, and to be as close to attractor models studied  in  \cite{Dalianis:2018frf}, we had better ignore this term. This can be easily achieved if the function $ f$ in Eq.(\ref{dal2}), which is arbitrary,   includes a compensating factor to cancel the  $cosh^{-6a}\left( \frac{\varphi}{\sqrt{6 \alpha}} \right) $ term.
 With its omission the potential in Eq.(\ref{dalfin}) resembles the one used in  \cite{Dalianis:2018frf} with the addition of only the last term which is proportional to $m_1$.  This model has a better chance to agree with data since for extremely low values of $m_1$ the model is exactly that of  \cite{Dalianis:2018frf} . 
 
\begin{figure}[t]
\centering
  \centering     
 \includegraphics[width=0.49\linewidth]{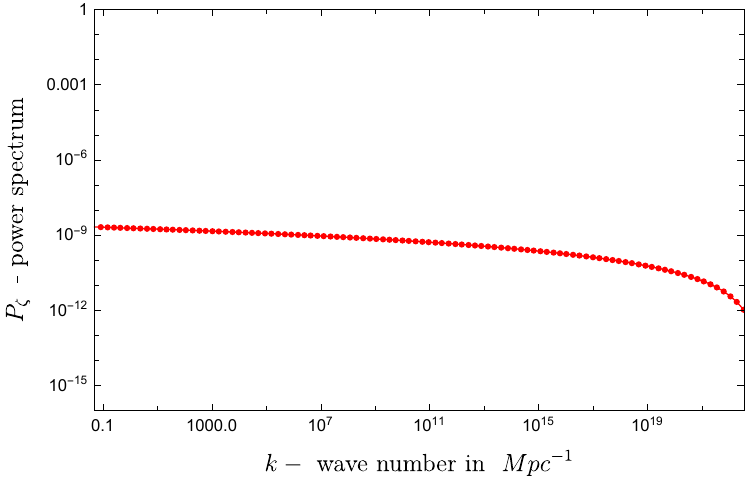}   
   \includegraphics[width=0.49\linewidth]{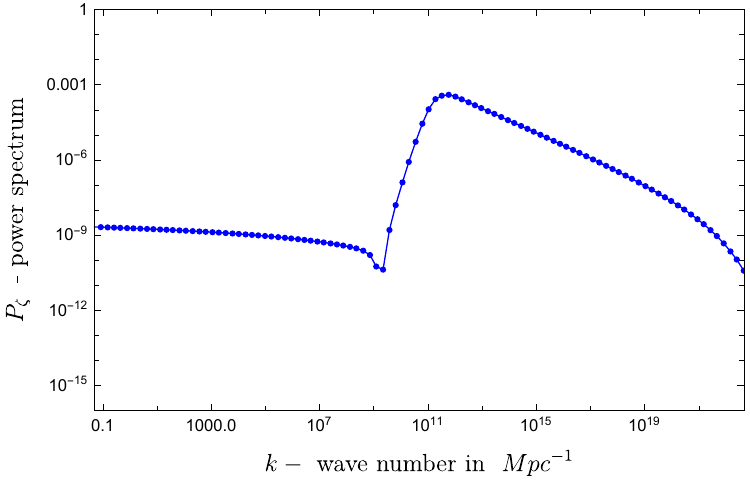}     
\caption{
The power spectrum for the modified $\alpha$-attractor model discussed in the main text. On the left $ b_1 = b_3 = 1$ and $m_1 = 0.01$ and 
no USR phase is present. On the right $ b_3 = -2.75, b_1 =  0.661713,  m_1 = 0.199575 $ and a USR phase exists.  In both cases 
$ \alpha =1 , b_0 = 0$. 
 }
   \label{figno}
\end{figure}

 With this modification we can obtain low values of $r$. For instance taking the parameters as $\alpha=1, b_1 = b_3 = 1$ and $m_1 = 0.01$ we get 
 \bea
\eta_s = 0.9646\,, \quad  \quad r = 0.0041  \,   .
 \eea  
 Thus both the spectral index and the tensor to scalar ratio are comfortably within their allowed limits.  The scale $V_0$ is taken $ V_0 = 1.47 \times 10^{-11}$  leading to 
 $ P_{0.05} \simeq 2.12 \times 10^{-9}   $.  This case does not yield any enhancement for the power spectrum since no  USR phase is present during the inflationary era. This is displayed in 
 the left pane of   Figure \ref{figno} where one observes a smooth behaviour of the power spectrum. 
 
 Cases where an USR phase is present giving rise to an enhancement of the power spectrum are feasible, by properly tuning the parameters.  
  In particular for $\alpha = 1$ and for values 
$ b_0 =0, b_3 = -2.75 $ and taking $b_1 =  0.661713,  m_1 = 0.199575 $ an inflection point exists at  $\varphi_{inf} =1.6 $,  with $\delta = - 0.005$,  giving rise to a USR evolution during inflation.  The predictions in this case are 
\bea
\eta_s = 0.9636\,, \quad  \quad r = 0.056  \,   ,
 \eea
 and the scale is  $V_0 = 7.78 \times 10^{-10}$ resulting to $ P_{0.05} = 2.08 \times 10^{-9}$.  The power spectrum has a maximum $ P_{max} = 3.91 \times 10^{-4}$ at 
 $k_{max} \simeq 5.9 \times 10^{11} \, Mpc^{-1}$ as shown on the right pane of  Figure \ref{figno}.  
 
The conclusion, concerning the modified $\alpha$-attractor scheme,  is that it can be in agreement with cosmological data, but hard to reconcile with large values for the power spectrum, pertinent to the creation of primordial black holes, with low tensor to scalar ratio $r \lesssim 0.04$ values as recent analyses impose.

\section{Conclusions}

We explore supergravity inflationary models driven by  supersymmetry breaking of a five dimensional supergravity, compactified on a $ S_1/Z_2 $ orbifold.
 The supersymmetry breaking takes place on the hidden brane and is transmitted to the visible brane by the radion field giving rise to an inflaton potential in which the quadratic and trilinear terms are gravitationally generated.  This mechanism yields inflaton potentials that differ from those studied in the literature. Within this framework we embed models having the structure of the no-scale form or $\alpha$- attractors and discuss their cosmological predictions.

We found that in the case of the no-scale model we have acceptable cosmological evolution as far as the tensor to scalar ratio and $\eta_s $ are concerned but this evolution can hardly account for PBH production mainly due to the requirement of  higher $r$ for significant enhancement of the specrtum. In the case of   $\alpha$- attractors we have not achieved to find  tensor to scalar ratio $r \lesssim 0.07$ even if we do not seek for PBH production. Certainly our findings can be improved significantly modifying either  the K\"{a}hler  function $\, K( \phi_i, \phi^*_i) $  or the superpotential $W$ as is shown in the last part of the previous section. Nevertheless such modifications are not easily seen how to emerge from a more fundamental point of view. 
The question whether the inclusion of higher orders in our approximation,  concerning $V_{brane}$ and the gravitationally induced brane to brane mediation of supersymmetry breaking which also affects the potential, will improve the situation remains to be seen.  This will be studied in a future work.

\newpage

\bibliography{orbiX}

\end{document}